# Detection of biomolecules and bioconjugates by monitoring rotated grating-coupled surface plasmon resonance

ANIKÓ SZALAI[1], EMESE TÓTH[1], ANIKÓ SOMOGYI[1], BALÁZS BÁNHELYI[2], EDIT CSAPÓ[3], IMRE DÉKÁNY[3], TIBOR CSENDES[2], MÁRIA CSETE[1*]

[1]*Department of Optics and Quantum Electronics, University of Szeged H-6720, Dóm tér 9, Szeged, Hungary*
[2]*Institute of Informatics, University of Szeged, H-6720, Árpád tér 2, Szeged, Hungary*
[3]*MTA-SZTE Supramolecular and Nanostructured Materials Research Group, Department of Medical Chemistry, Faculty of Medicine, University of Szeged, H-6720, Dóm tér 8, Szeged, Hungary*
*mcsete@physx.u-szeged.hu

**Abstract:** Plasmonic biosensing chips were prepared by fabricating wavelength-scaled dielectric-metal interfacial gratings on polymer film covered bimetal layers. Lysozyme biomolecules (LYZ) and gold nanoparticle bioconjugates (AuNP-LYZ) with 1:5 mass ratio were seeded onto the biochip surfaces. Comparison of the reflectance curves measured in a modified Kretschmann arrangement and computed numerically proved that monitoring the narrower secondary peaks under optimal rotated-grating coupling condition makes it possible to achieve enhanced sensitivity in biodetection. The enlarged resonance peak shift is due to the horizontally and vertically antisymmetric long-range plamonic modes propagating at the edge of the valleys and hills, which originate from Bragg scattered surface plasmon polaritons. The sensitivity is further increased in case of bioconjugates due to the coupled localized resonances on Au NPs.

_________________________________________________________________

**References and links**:
(1) Elson, J. M. Light scattering from surfaces with a single dielectric overlayer. *J. Opt. Soc. Am.*, **66 (7)**, 682-694 (1976)**.**
(2) Giannattasio, A.; Hooper, I. R.; Barnes, W. L. Dependence on surface profile in grating-assisted coupling of light to surface plasmon-polaritons. *Opt. Comm.*, **261**, 291-295 (2006)**.**
(3) Wang, B.; Lalanne, P. Surface plasmon polaritons locally excited on the ridges of metallic gratings. *J. Opt. Soc. Am.*, **27/6**, 1432-1441 (2010)**.**
(4) Bozhevolnyi, S. I.; Sonergaard T. General properties of slow-plasmon resonant nanostructures: nano-antennas and resonators. *Optics Express*, **15**(17), 10869 (2007).
(5) D'Aguanno G.; Mattiucci, N.; Alú, A.; Bloemer, M. J. Quenched optical transmission in ultrathin subwavelength plasmonic gratings. *Phys. Rev B*, **83**, 035426 (2011).
(6) Chen, Y. J.; Koteles, E. S.; Seymour, R. J.; Sonek, G. J.; Ballantyne, J. M. Surface plasmons on gratings: coupling in the minigap regions. *Solid State Comm.,* , **46**(2), 95-99 (1983).
(7) Weber, M. G.; Mills, D. L. Determination of surface-polariton minigaps on grating structures: A comparison between constant-frequency and constant-angle scans. *Phys. Rev. B*, **34**(4), 2893-2894 (1986).
(8) Barnes, W. L.; Preist, T. W.; Kitson, S. C.; Sambles, J. R.; Cotter, N. P. K.; Nash, D. J. Photonic gaps in the dispersion of surface plasmons on gratings. *Phys. Rev. B*, **51**(16), 11164–11167 (1995).

(9) Ameling, R.; Giessen, H. Cavity Plasmonics: Large Normal Mode Splitting of Electric and Magnetic Particle plasmons Induced by a Photonic Microcavity. *Nano Letters*, **10**, 4394 (2010).

(10) Ghoshal, A.; Divliansky, I.; Kik, P. G. Experimental observation of mode-selective anticrossing in surface-plasmon coupled metal nanoparticle arrays. *Appl. Phys. Lett.*, **94**, 171108 (2009).

(11) Mills, D. L. Interaction of surface polaritons with periodic surface structures; Rayleigh waves and gratings. *Phys. Rev. B*, **15**, 3097–3118 (1977).

(12) Seshadri, S. R. Coupling of surface polaritons incident obliquely on a small amplitude grating. *J. Appl. Phys.*, **58**(5), 1733-1738 (1985).

(13) Hibbins, A. P.; Sambles, J. R.; Lawrence, C. R. Azimuth-angle-dependent reflectivity data from metallic gratings. *Journal of Modern Optics*, **45**(5), 1019-1028 (1998).

(14) Kretschmann, M.; Leskova, A.; Maradudin, A. A. Conical propagation of a surface plasmon polariton across a grating. *Optics Communications*, **215**, 205-223 (2003).

(15) Romanato, F.; Hong, L. K.; Kang, H. K.; Wong, C. C.; Yun, Z.; Knoll, W. Azimuthal dispersion and energy mode condensation of grating-coupled surface plasmon polaritons. *Phys. Rev. B*, **77**, 245435 (2008).

(16) Csete, M.; Vass, Cs.; Kokavecz, J.; Goncalves, M.; Megyesi, V.; Bor, Zs.; Pietralla, M.; Marti, O. Effect of sub-micrometer polymer gratings generated by two-beam interference on surface plasmon resonance. *Appl. Surf. Sci.*, **247**(1), 477-485 (2005).

(17) Csete, M.; Szekeres, G.; Vass, Cs.; Maghelli, N.; Osvay, K.; Bor, Zs.; Pietralla, M.; Marti, O. Surface plasmon resonance spectroscopy on rotated sub-micrometer polymer gratings generated by UV laser based two-beam interference. *Appl. Surf. Sci.*, **252**(13), 4773-4780 (2006).

(18) Csete, M.; Kőházi-Kis, A.; Vass, Cs.; Sipos, Á.; Szekeres, G.; Deli, M.; Osvay, K.; Bor, Zs. Atomic force microscopical and surface plasmon resonance spectroscopical investigation of sub-micrometer metal gratings generated by UV laser based two-beam interference in Au-Ag bimetallic layers. *Appl. Surf. Sci.*, **253**, 7662-7671 (2007).

(19) Szalai, A.; Szekeres, G.; Balázs, J.; Somogyi, A.; Csete, M. Rotated grating coupled surface plasmon resonance on wavelength-scaled shallow rectangular gratings. *Proc. SPIE 8809, Plasmonics: Metallic Nanostructures and Their Optical Properties XI*, 88092U

(20) Parisi, G.; Zilio, P.; Romanato, F. Complex Bloch-modes calculation of plasmonic crystal slabs by means of finite element method. *Optics Express*, **20**(15), 16690-16703 (2010).

(21) Gazzola, E.; Brigo, L.; Zacco, G.; Zilio, P.; Ruffato, P.; Brusatin, G.; Romanato, F. Coupled SPP Modes on 1D Plasmonic Gratings in Conical Mounting. *Plasmonics*, **9**, 867 (2013).

(22) Kim, D. Effect of the azimuthal orientation on the performance of grating-coupled surface-plasmon resonance biosensors. *Appl. Opt.*, **44**(16), 3218-3223 (2005).

(23) Csete, M.; Kőházi-Kis, A.; Megyesi, V.; Osvay, K.; Bor, Zs.; Pietralla, M.; Marti, O. Coupled surface plasmon resonance on bimetallic films covered by sub-micrometer polymer gratings. *Org. Electronics*, **8**(2), 148-160 (2007).

(24) Csete, M.; Sipos, Á.; Kőházi-Kis, A.; Szalai, A.; Szekeres, G.; Mathesz, A.; Csákó, T.; Osvay, K.; Bor, Zs.; Penke, B.; Deli, M. A.; Veszelka, Sz.; Schmatulla, A.; Marti, O. Comparative study of sub-micrometer polymeric structures: dot-arrays, linear and crossed gratings generated by UV laser based two-beam interference, as surfaces for SPR and AFM based bio-sensing. *Appl. Surf. Sci.*, **254**(4), 1194-1205 (2007).

(25) Tóháti, H.; Sipos, Á.; Szekeres, G.; Mathesz, A.; Szalai, A.; Jójárt, P.; Budai, J.; Vass, Cs.; Kőházi-Kis, A.; Csete, M.; Bor, Zs. Surface plasmon scattering on polymer-bimetal layer covered fused silica gratings generated by laser-induced backside wet etching. *Appl. Surf. Sci.*, **255**(10), 5130-5137 (2009).

(26) Sipos, Á.; Tóháti, H.; Mathesz, A.; Szalai, A.; Veszelka, Sz.; Deli, M. A.; Fülöp, L.; Kőházi-Kis, A.; Csete, M.; Bor, Zs. Effect of nanogold particles on coupled plasmon resonance on biomolecule covered prepatterned multilayers. *Sensor Lett.*, **8**, 512-520 (2010).

(27) Romanato, F.; Lee, K. H.; Kang, H. K.; Ruffato G.; Wong, C. C. Sensitivity enhancement in grating coupled surface plasmon resonance by azimuthal control *Optics Express*, **17(14),** 12145-12154 (2009).

(28) Brigo1, L.; Gazzola, E.; Cittadini, M.; Zilio, P.; Zacco, G.; Romanato, F.; Martucci, A.; Guglielmi, M.; Brusatin, G. Short and long range surface polariton waveguides for xylene sensing. *Nanotechnology*, **24**, 155502 (2013).

(29) González, M. U.; Weeber, J.-C.; Baudrion, A.-L.; Dereux, A.; Stepanov, A. L.; Krenn, J. R.; Devaux, E.; Ebessen, T. W.; Design, near-field characterization, and modeling of 45° surface-plasmon Bragg mirrors. *Phys. Rev. B*, **73**, 155416 (2006).

(30) Randhawa, S.; González, M. U.; Renger, J.; Enoch, S.; Quidant, R. Design and properties of dielectric surface plasmon Bragg mirrors. *Opt. Exp.*, **18**(14), 14496-14510 (2010).

(31) Homola, J.; Yee, S. S.; Gauglitz, G. Surface plasmon resonance sensors: review. *Sens. Actuators B*, **54**, 3-15 (1999).

(32) Kim, D.; Shuler, M. L. Design and development of grating coupled optical biosensor to detect animal pathogen. *Proc of SPIE*, **5321**, Biomedical Vibrational Spectroscopy and Biohazard Detection Technologies, 309-314 (2004).

(33) Lukosz, W. Integrated optical chemical and direct biochemical sensors. *Sens. Actuators B*, **29**. 37-50 (1995).

(34) Ramsden, J. J.; Optical Biosensors. *J. Mol. Recognit.*, **10**, 109-120 (1997).

(35) Brockman, J. M.; Fernandez, S. M. Grating-coupled surface plasmon resonance for rapid, label-free, array-based sensing. *Am. Lab.*, **33**. 37-40 (2001).

(36) Zieziulewicz, T. J.; Unfricht, D. W.; Hadjout, N.; Lynes, M. A.; Lawrence, D. A. Shrinking the biologic world – nanobiotechnologies for toxicology. *Toxicol. Sci.*, **74**. 235-244 (2003).
(37) Senlik, S. S.; Kocabas, A.; Aydinli, A.; Grating based plasmonic band gap cavities. *Opt. Exp.*, **17**(18), 15541-15549 (2009).
(38) Wang B.; Wang, G. P. Plasmon Bragg reflectors and nanocavities on flat metallic surfaces. *Appl. Phys. Lett.*, **87**, 013107 (2005).
(39) Shi, H.; Liu, Z.; Wang, X.; Guo, J.; Liu, L.; Luo, L.; Guo, J.; Ma, H.; Sun, S.; He, Y. A symmetrical optical waveguide based surface plasmon resonance biosensing system. *Sensors and Actuators B*, **185**, 91 (2013).
(40) Lal, S.; Link, S.; Halas, N. J. Nano-optics from sensing to waveguiding. *Nature Photonics*, **1**, 641-648 (2007).
(41) Anker, J. N.; Hall, W. P.; Lyandres, O.; Shah, N. C.; Zhao, J.; Van Duyne, R. P. Bio-sensing with plasmonic nanosensors. *Nature*, **7**, 442-453 (2007).
(42) Luk'yanchuk, B.; Zheludev, N. I.; Maier, S. A.; Halas, N. J.; Nordlander, P.; Giessen, H.; C. T. Chong *Nature Materials*, **9**, 707-715 (2010).
(43) Formoso, C.; Forster, L. S. Tryptophan fluorescence lifetimes in lysozyme. *Journal of Biological Chemistry*, **250**, 3738-3745 (1975).
(44) Wei, H.; Wang, Z.; Yang, L.; Tian, S.; Hou, C.; Lu, Y. Lysozyme-stabilized gold fluorescent cluster: Synthesis and application as Hg2+ sensor, Analyst., **135**, 1406-1410 (2010).
(45) Chen, W. Y.; Lin, J. Y.; Chen, W. J.; Luo, L.; Diau, E.W.G.; Chen, Y.C. Functional gold nanoclusters as antimicrobal agents for antibiotic-resistant bacteria. *Nanomedicine*, **5**, 755-764 (2010).
(46) Hornok, V.; Csapó, E.; Varga, N.; Ungor, D.; Sebők, D.; Janovák, L.; Laczkó, G.; Dékány I. Controlled syntheses and structural characterization of plasmonic and red-emitting gold/lysozyme nanohybrid dispersions. *Collid. Polym. Sci.*, **294**, 49-58 (2015).
(47) Arwin, H. Optical properties of thin layers of bovine serum albumin, γ-globulin, and hemoglobin. *Applied Spectroscopy*, **40**(3), 313-318 (1986).
(48) Johnson P. B.; Christy R. W. Optical constants of the noble metals. *Phys. Rev. B*, **6**(12), 4370-4379 (1972).
(49) Derjaguin B. V.; Muller V. M.; Toporov Yu. P. On the role of molecular forces in contact deformations. *Colloid Interf. Sci.*, **53**, 378 (1978).
(50) Csete M.; Kurdi G.; Kokavecz J.; Megyesi V.; Osvay K.; Schay Z.; Bor Zs.; Marti O. Application possibilities and chemical origin of sub-micrometer adhesion modulation on polymer gratings produced by UV laser illumination. Mat. Sci. and Engin. C **26** 1056-1062 (2006).

___________________________________________________________________________

## 1. Introduction:

Surface plasmon polaritons (SPPs) propagating either on flat or on patterned surfaces have intriguing far- and near-field properties. In the primary literature of SPPs there are several papers describing the effects, which are exerted on SPR by different periodic surface corrugations [1]. The surface profile, the period and the modulation amplitude together determines the coupling efficiency of light into SPPs, which is maximal, when the fill-factor is 50% [2]. Detailed theoretical studies performed to analyze SPP excitation efficiency on periodic slit and groove arrays on metallic surfaces have shown that the SPP excitation efficiency is the highest close to the Rayleigh condition [3]. Characteristic differences between optical responses originating from short-range (SRSPP) and long-range (LRSPP) plasmonic modes as well as from localized surface plasmon resonances (LSPR) involved into the grating-coupling phenomena were also uncovered [4, 5].

Presence of gratings results in opening of band-gaps in those parameter regions, where Bragg scattering occurs [6-10]. Caused by possible misinterpretation of extrema at the minigaps, it was concluded that to uncover plasmonic band gaps (PBG) a more reliable method is to plot the extrema as a function of frequency with a fixed angle of incidence [6, 7]. The physical origin of PBGs was ascribed to the coexistence of two modes, which possess different field and charge distributions [8]. Band gap opening on the dispersion characteristics can originate from strong-coupling between localized and propagating modes as well. Besides the overlapping frequencies corresponding to resonances, necessary condition of strong - coupling is the appropriate symmetry of interacting localized and propagating modes [9]. Coupling between symmetric and antisymmetric modes on nanowires and cavity resonances manifests itself in a large anticrossing, which provides a general design-scheme for three-dimensional Bragg structures' design [10].

Significantly smaller amount of theoretical and experimental studies has been published about the effect of periodic structures' azimuthal orientation [11-21]. These studies resulted in development of several novel biosensing methodologies [22-28]. Azimuthal orientation dependent grating-coupling of evanescent and radiative modes was investigated on sinusoidal gratings [13]. It was proven that the overlapping bands of modes supported by structured multilayers manifests themselves in a gap at the Brillouin zone boundary [11, 12, 14]. In most of grating studies preformed in a conical mounting the light-to-SPP coupling is realized by the periodic structure directly [15, 20-22, 27, 28]. It was shown that two SPPs possessing the same wave vector, but different propagation directions can be excited in the azimuthal orientations corresponding to the so-called grating-coupled surface plasmon resonance (GCSPR) [15, 20, 21, 28]. In our previous studies we have presented another phenomenon, the rotated grating-coupling surface plasmon resonance (RGC-SPR) occurring, when SPP excitation is performed on a periodically structured multilayer, which is aligned in a conical mounting onto a prism inside a modified Kretschmann setup [16-19, 23-26]. Two SPPs with different wave vectors are excitable inside a narrow azimuthal orientation region, which result in double minima on the polar angle dependent reflectance [19].

Periodic metallic surfaces are important elements in design of future plasmonic devices, such as SPP generators, couplers and reflectors [29]. Dielectric ridges on metal films in proper azimuthal orientation act as Bragg mirrors due to the large reflectance inside the gap [30]. In recent nanophotonics there are tremendous efforts to overcome the limitations of conventional biosensors via SPR phenomena. Traditional SPR spectroscopy (SPRS) methods are based on the sensitivity of propagating SPPs' resonance characteristics to the dielectric environment [31]. Metal gratings were primarily applied as in-coupling elements in waveguide based SPR optical biosensors [32-34]. Important application area of plasmon Bragg-gratings is biosensing, due to the enhanced sensitivity of grating-coupled modes [35, 36]. High sensitivity intra-cavity biosensing has been realized via cavity modes arising inside the plasmonic band gaps on selectively loaded gratings and via defects in waveguides [37, 38].

Application of grating-coupled SPR biosensors relying on multiple diffracted order's monitoring at different azimuthal orientations is limited by the rotation sensitivity of the conical mounting [22]. However, it was demonstrated that enhanced sensitivity is achievable by monitoring the reflectance peaks originating from the two identical SPPs in GCSPR configuration [27]. The advantages of long-range modes, manifesting themselves in narrower peaks and accompanied by higher and spatially broader **E**-field enhancement, were also considered [21, 28, 39]. In our previous experimental studies, we have shown that rotation into appropriate azimuthal orientation capable of resulting in double resonance peaks also makes it possible to enhance sensitivity in RGC-SPR based biosensing [23-26].

Novel class of localized surface plasmon resonance (LSPRS) sensing methods relies on the sensitivity of LSPR to the presence of dielectric cover layers on plasmonic nano-objects [40, 41, 42]. At the same time noble metal particles can play multiple role, since they can enhance the local **E**-field intensity, accordingly can be used as a kind of markers on SPP-based bioplatforms, moreover can initiate cavity modes under certain circumstances.

While preparing a plasmonic biosensor for a specific molecule, it has to be ensured that the **E**-field enhancement is strong at those locations, where the molecule prefers to attach. Local EM-field enhancement at those wavelengths, where fluorescent molecules absorb and emit light, makes it possible to achieve higher sensitivity in detection via SPR enhanced fluorescence phenomena. Several intriguing biomolecules are fluorescent, among them lysozyme (LYZ) is an important protein having a potential to use in the field of medicine due to its antibacterial capabilities. The fluorescence of LYZ originates from tryptophan, and sensitively depends on the chemical and dielectric environment [43].

Various noble metal nanoclusters (NCs) and aggregates are applied in bioimaging due to their optical properties, non-toxicity, stability and solvability in water. The formation of biocompatible gold and silver NCs can be promoted by adding proteins, e.g. LYZ into the aqueous solution of $AuCl_4^-$ precursor and the spontaneous interaction of LYZ with the aurate ions results in the formation of LYZ-stabilized Au NCs or Au NPs depending on the applied mass ratio of the reactants. Fluorescence of LYZ in clusters and aggregates is retained, moreover, Au-LYZ complexes are sensitive to Hg concentration, which can be used in detectors [44]. In addition to this, ~1 nm sized Au NCs as well as ~10 nm sized Au NPs coated by LYZ possess antimicrobial capabilities [45]. Recent studies revealed that Au:LYZ 1:5 mass ratio results in a protein shell around NP core type bioconjugate, while in case of smaller ratios, Au NC seeds are distributed in protein islands [46]. Tuning SPP or LSPR phenomena into the excitation and emission bands, which strongly depend on bioconjugates composition is challenging, as a consequence there is a great demand for alternate bio-sensing methodologies.

The purpose of our present work was to demonstrate the enhanced sensitivity, which is achievable by monitoring the resonance peaks arising due to rotated grating-coupling phenomenon in narrow azimuthal orientation regions. Another purpose was to answer the question, whether sensitivity to LYZ protein can be further increased by using AuNP-LYZ bioconjugates in RGC-SPR. Experimental reflectance curves from angle interrogation of RGC-SPR phenomenon were compared to resonance curves determined by theoretical computations. The accompanying near-field distribution was inspected to uncover the underlying nanophotonical phenomena. Further computations were performed to demonstrate the existence of a configuration optimal to achieve maximal polar angle shift, and to map the dispersion characteristics of the responsible grating-coupled SPPs. As a result, a general methodology is proposed, which makes it possible to achieve enhanced sensitivity on grating-based sensing chips in an optimal conical mounting.

## 2. Materials and Methods:

### 2.1. Experimental methods

#### 2.1.1. Preparation of plasmonic AuNP:LYZ nanodispersion

For the preparation of LYZ reduced Au NPs, LYZ (≥90%, Sigma-Aldrich), $HAuCl_4 \cdot 3H_2O$ (≥ 49.0 %, Sigma-Aldrich) and sodium hydroxide (99 %, Molar Chemicals) were used. During preparation LYZ solution with 0.98 mg/ml concentration and 10 ml of $HAuCl_4$ solution with 1.0 mM concentration were admixed. The appropriate pH 12 was adjusted with 2 M NaOH solution. After 18 h incubation at 40 °C the color modification of the solution indicated the Au NPs formation. Different AuNP-LYZ bioconjugates were synthetized and the diameter of NPs was altered by tuning the fraction of LYZ. Dispersions prepared with a ratio of $m_{Au}:m_{LYZ}$ = 1:5 became red color, indicating LSPR related absorptance close to the 532 nm wavelength of SPR interrogation. Accordingly, this dispersion was selected for RGC-SPR sensing of bioconjugates.

#### 2.1.2. Characterization of AuNP-LYZ bioconjugates

Transmission electron microscopy (TEM) was used to visualize the Au NPs with Technai (200 kV) apparatus/tool and then the size of bioconjugates was estimated based on the TEM images (UTHSCSA Image Tool 2.00 software) (Fig. 2a, inset). The measurement of the absorption spectrum was carried out with UVIKON 930 Type dual-beam spectrophotometer. The emission spectrum was measured with Horiba Jobin Yvon Fluoromax-4 spectro-fluorometer with 365 nm excitation wavelength and the value of the slit was 3 nm (Fig. 2a).

#### 2.1.3. Experimental investigation of RGC-SPR

For biochip preparation the NBK7 glass substrates were evaporated by 38 nm thick silver and 7 nm thick gold films. Bimetal layer is necessary to realize efficient plasmon coupling, since silver has good plasmonic properties, while the thin gold cover-layer protect the silver from corrosion [16-19, 23-26]. On the bimetal surface a ~60 nm thick polycarbonate (PC) layer was spin-coated, and one dimensional gratings with a sinusoidal profile were fabricated via two-beam interference laser procedure (Fig. 1a-d). In our previous studies it was shown that rotated grating-coupling of SPPs at 532 nm excitation occurs on the wavelength-scaled 416 nm periodic PC grating covered bimetal layer, when the *a* modulation amplitude is larger than a minimal value determined by the multilayer composition [23]. The thinnest average polymer layer is 15.75 nm, which makes it possible to achieve high coupling efficiency on a grating with 2*a*>31.5 nm modulation depth [16-19, 23-26].

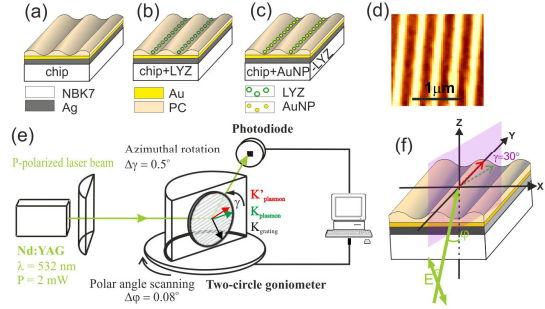

Fig. 1. Schematic drawings of the investigated biosensor chips: (a) sinusoidal polymer grating, (b) sinusoidal grating covered by LYZ shells and (c) sinusoidal grating covered by AuNP-LYZ bioconjugates, (d) AFM picture of the grating surface. (e) Schematic drawing of the SPR setup based on a modified Kretschmann arrangement, used to study the RGC-SPR phenomenon. (f) Method of polar and azimuthal angle tuning.

The 10 mJ/cm$^2$ laser fluence applied in chip preparation ensured that the polymer grating modulation depth was approximately double of the minimal value, as a result the rotated grating-coupling condition was met. For RGC-SPR based biodetection measurements LYZ biomolecule and AuNP-LYZ bioconjugate containing solutions were seeded onto the structured multilayers (Fig. 1a-c). The angle interrogation of the RGC-SPR was performed in a modified Kretschmann-like arrangement by varying the polar angle in $\Delta\varphi = 0.08°$ steps (Fig. 1e). The 532 nm p-polarized light was coupled into SPPs ($K_{plasmon}$) via a half-cylinder, while the efficient grating-coupled SPP mode ($K'_{plasmon}$) excitation was ensured by the azimuthal orientation of the grating. The azimuthal orientation of the grating grooves with respect to the plane of incidence was set to $\gamma \sim 30°$ via special holders, which ensured rotation with +/-0.5 ° accuracy (Fig. 1f).

A frequency doubled Nd:YAG laser (Intelite, GSLN32-20, $\lambda$ =532 nm, 2 mW) was used to realize SPP excitation. The reflected light was collected by a standard visible range photodetector (Thorlabs DET 110). To ensure the appropriate $\varphi$-$2\varphi$ rotation of the half-cylinder and the photodetector, a two-circle goniometer (OWIS, with DMT 65, 2-Ph-SM 240) was used. The RGC-SPR phenomenon was angle interrogated first on bare chips for reference purposes (Fig. 1a), then on a sensor-chip covered by LYZ (Fig. 1b), and finally the effect of AuNP-LYZ bioconjugates (Fig. 1c) was studied. The measured reflectance curves are presented in Fig. 2b.

## 2.2. Theoretical methods

Numerical computations were carried out with COMSOL Multiphysics software package (COMSOL AB) based on Finite Element Method (FEM). Radio Frequency Module was used to calculate reflectance from the structured multilayer in a conical mounting. The electromagnetic near-field distribution was also inspected at the resonance peaks on reflectance in order to uncover the nature of SPP modes that are at play in grating-coupled resonance. To determine the symmetry properties of SPPs coupled in specific illumination configurations, both the normalized **E**-field and the $E_y$ field component were studied on the horizontal and vertical plane cross-sections of the inspected unit cells. The vertical plane cross-section along the valleys was taken at the turning line of the $E_y$ field component, which is the longitudinal component for modes propagating along the stripes.

The wavelength-dependent optical properties of all components were taken into account. The refractive index of LYZ was considered by a general Cauchy-formula of proteins ($n_{LYZ} = AC + BC/\lambda^2$, where $AC$=1.45, $BC$=0.01m$^2$) according to the literature [47]. The wavelength-dependent complex dielectric functions of Au and Ag layers were implemented based on the literature, by interpolating the measured data set with spline-fits [48].

### 2.2.1. Numerical modeling of RGC-SPR on a fitted chip

The modeling of a fitted chip has been performed to analyze the effect of the experimental conditions (multilayer composition, illumination direction, protein location) on the reflectance and to uncover the physical origin of the coupled resonance peaks arising due to RGC-SPR by analyzing the accompanying near-field phenomena (Fig. 2c, 3). The fitted chip is an artificial multilayer, with a fitted sinusoidal grating profile and azimuthal orientation, which results in resonance peaks approximating well the measurements. First the reflectance curves measured on bare chips were fitted by supposing the existence of a sinusoidal PC grating and by varying the layer thickness at the minima (bottom of the valleys) and at the maxima (top of the hills) of the sinusoidal modulation, as well as by tuning the azimuthal orientation (Fig. 1a, 2b-to-c). Then the reflectance curve measured on a LYZ covered chip was fitted, by varying the dielectric layer thickness only in the grating valleys. This is in accordance with theoretical computations and AFM measurements, which prove the existence of a sinusoidal adhesion modulation of topography origin [49, 50]. The co-existent sinusoidal adhesion modulation promotes the adherence of biomolecules and bioconjugates inside the valleys of the polymer grating [24, 25, 27, 50]. The equivalent protein layer thickness was computed by taking into account the different dielectric properties of the PC and LYZ.

The composition of biomolecule layers and bioconjugates in the COMSOL models was simplified. The 13.4 nm AuNP diameter was selected based on the size distribution measured on TEM images (Fig. 2a inset). The 1.8 nm thickness of LYZ shell was selected based on the thickness of monomolecular covering. Accordingly, linear arrays of empty LYZ shells corresponding to the previously fitted protein layer thickness were inserted into the grating valleys (Fig. 1b). Namely, dielectric shells with 6.7 nm and 8.5 nm inner and outer radii were aligned along the center of the valleys. By taking into account that the gold NPs presence may alter the number of biomolecules adhered from a dispersion of specific concentration, such a different number of core-shell particles consisting of Au NPs covered by LYZ shells was selected, which ensured a resonance peak shift corresponding to the measurement on the bioconjugates covered chip (Fig. 1c). In our previous studies an adhesion enhancement of non-topographical origin was demonstrated at the edge of the valleys and hills [50]. Accordingly, both the LYZ shells and AuNP-LYZ bioconjugates were relocated from the bottom to the hillside, to inspect the existence of a preferred location, where the sensitivity can be further enhanced (Fig. 2c, 3). The exact location on the hillside was selected taking into account the distribution of the normalized **E**-field and the $E_y$ field component, as it is described in **Sections 3.3 and 3.5**.

### 2.2.2. Numerical modeling of RGC-SPR on a designed chip

Detailed theoretical study of a designed biochip has been performed as well to demonstrate the RGC-SPR phenomenon in an optimal configuration (Fig. 2d-f, 4-6). The designed chip is an artificial multilayer consisting of a sinusoidal grating profile approximating the minimal average polymer layer thickness and modulation depth capable of resulting in RGC-SPR phenomenon in case of the applied PC on Ag-Au bimetal multilayer composition on NBK7 substrate. There was an additional thin polymer layer in the valleys to avoid problems with meshing. The reflectance was studied on different multilayers by varying the polar angle in [38°, 78°] interval and by tuning the azimuthal orientation in [28°, 38°] interval, both with 1° steps (Fig. 2d-f and 4). The reflectance curves of bare chips, and biochips covered by LYZ and AuNP-LYZ bioconjugate were compared for three different azimuthal orientations to prove the existence of an optimal orientation (Fig. 2d-f). In general, the optimal orientation of a chip with a specific multilayer composition is qualified by the azimuthal angle, which makes it possible to achieve the maximal polar angle shift.

To analyze the overlap between the coupled modes and bio-objects, the normalized **E**-field and the $E_y$ field component were studied in the horizontal and vertical planes of two unit cells of the designed chip (Fig. 5, with corresponding Visualization 1 provided in Supporting Information). The purpose was to show that the optimal configuration makes it possible to improve the spatial overlap between the grating-coupled LRSPPs and bio-objects at the bottom of the valleys, where the adhesion modulation inherently promotes their adherence [24, 25, 27, 50].

In order to explain the nanophotonical origin of the coupled resonance peaks, the dispersion characteristics of bare, LYZ and AuNP-LYZ bioconjugate coated designed biochips were determined as well (Fig. 6). The wavelength was swept in [300 nm, 700 nm] interval with 10 nm steps, while the polar angle was swept in [38°, 78°] interval with 1° steps.

## 3. Results and Discussion

Based on TEM measurements the average diameter of AuNP-LYZ 1:5 particles was ~13.4 nm. These dispersions exhibit a fluorescence with an emission maximum at 450 nm in case of excitation at 365 nm. The spectral study revealed that the dispersion with $m_{Au}:m_{LYZ}$ = 1:5 mass ratio exhibits a well-defined absorption at 529 nm, which is close to the 532 nm wavelength used for SPP excitation (Fig. 2a). Accordingly, all SPP coupling phenomena, which result in EM-field enhancement at this wavelength, promote the absorption and mediately increase the sensitivity to the bioconjugates. As a result, the fluorescent bio-conjugates can be detected also via an illumination outside their excitation and emission bands.

### 3.1. Experimental demonstration of RGC-SPR

The experimental RGC-SPR studies revealed that double peaked reflectance curves are measurable in $\gamma \sim 30°$ azimuthal orientation on multi-layers consisting of 416 nm periodic gratings with $2a_{\text{chip}\_1} \sim 2a_{\text{chip}\_2} \sim 63.00$ nm modulation depth. Both the secondary and the primary peaks correspond to SPP modes, which propagate on the modulated grating surface. The applied nominations reveal that the location of the secondary / primary resonance peak is in a region of smaller polar angles / resembles to that originating from SPPs on a flat surface. All azimuthal and polar angles corresponding to resonance peaks are collected in Table 1. The secondary peaks appeared at $\varphi_{\text{chip}\_1}^{\sec ondary} = 52.16°$ and $\varphi_{\text{chip}\_2}^{\sec ondary} = 50.72°$ polar angles on the studied bare biochips (Fig. 2b).

Covering by LYZ causes a secondary peak at $\varphi_{\text{chip}\_1+LYZ}^{\sec ondary} = 52.48°$ polar angle, which corresponds to a moderate $\Delta\varphi_{\text{chip}\_1+LYZ}^{\sec ondary} = 0.32°$ shift in polar angle. Covering by AuNP-LYZ bioconjugate results in a secondary peak at $\varphi_{\text{chip}\_2+AuNP-LYZ}^{\sec ondary} = 51.6°$, which indicates a more than two-times larger $\Delta\varphi_{\text{chip}\_2+AuNP-LYZ}^{\sec ondary} = 0.88°$ polar angle shift. These results unambiguously prove that presence of Au NPs results in a significantly larger shift of the secondary peak. The measurements also indicate that the observation possibility of both peaks sensitively depends on the exact multilayer composition. Namely, at $\gamma \sim 30°$ azimuthal orientation the $\varphi_{\text{chip}\_1}^{primary} = 64.96°$ primary peak is bakward shited by $\Delta\varphi_{\text{chip}\_1+LYZ}^{primary} = -0.76°$ and flattened already, when only LYZ was adhered to the surface, while the secondary peak was still suitable for detection. In contrast, the primary peak at $\varphi_{chip\_2}^{primary} = 69.44°$ was preserved during the adherence of AuNP-LYZ conjugates, but it was shifted backward by $\Delta\varphi_{chip\_2+AuNP-LYZ}^{primary} = -4.8°$, and at the same time significantly broadened. These results indicate that the broader primary peaks are less suitable for detection.

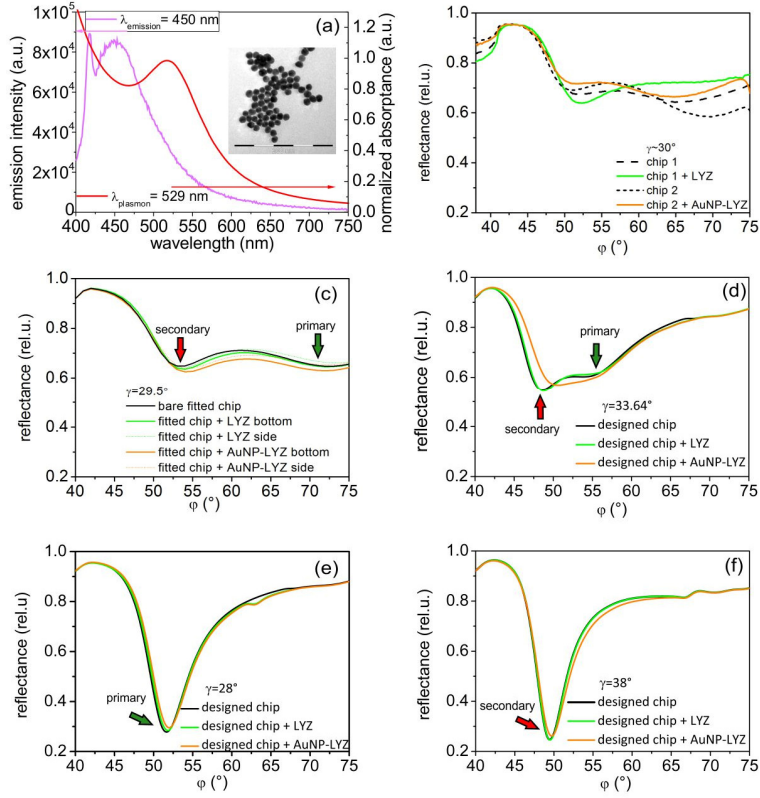

Fig. 2. (a) Absorptance and emission spectra of AuNP-LYZ nanodispersions with $m_{Au}$:$m_{LYZ}$ =1:5 mass ratio, the inset shows TEM image of the AuNP-LYZ bioconjugates. Reflectance before and after covering by LYZ and AuNP-LYZ bioconjugates (b) measured on two sensor chips, (c) computed on a fitted sensor chip in case of bio-objects at the bottom of valleys and at the side of the hills, (d-f) computed on the designed sensor chip in case of bio-objects at the bottom of valleys at different azimuthal orientations: (d) $\gamma = 33.64°$; (e) $\gamma = 28°$; (f) $\gamma = 38°$.

### 3.2. Modeling reflectance governed by RGC-SPR on a fitted chip

Based on the fitting of surface profiles of uncovered chips, the modulation amplitude on both multilayers is $2a_{fitted\_chip\_1}$~ $2a_{fitted\_chip\_2}$~ 63 nm, and there is no PC layer in the valleys. The fitted azimuthal orientation is $\gamma_{fitted\_chip}$ ~ 29.5°, which equals to the $\gamma_{multilayer\_optimal} = 29.3°$ optimal azimuthal orientation computed for the fitted multilayer based on reference [23].

By supposing that LYZ and AuNP-LYZ bioconjugate are seeded exactly at the bottom of the valleys, adherence of $N_{fitted\_chip+LYZ\_bottom} = 120$ and $N_{fitted\_chip+AuNP-LYZ\_bottom} = 16$ number of LYZ shells and AuNP-LYZ core-shell type nano-objects per one unit cell resulted in $\Delta\varphi_{fitted\_chip+LYZ\_bottom}^{secondary} = 0.4°$ and $\Delta\varphi_{fitted\_chip+AuNP-LYZ\_bottom}^{secondary} = 0.9°$ secondary peak shifts, which are slightly larger than the measured shifts. The same amount of biomolecules and bioconjugates per unit cell caused $\Delta\varphi_{fitted\_chip+LYZ\_bottom}^{primary} = -0.2°$ and $\Delta\varphi_{fitted\_chip+AuNP-LYZ\_bottom}^{primary} = -0.6°$ backward primary peak shift, respectively.

**Table 1.** Secondary and primary resonance minima on the measured and computed reflectance curves.

| | | $\varphi^{sec}(°)$ | $\varphi_{LYZ}^{sec}(°)$ | $\varphi_{AuNP\text{-}LYZ}^{sec}(°)$ | $\varphi^{pri}(°)$ | $\varphi_{LYZ}^{pri}(°)$ | $\varphi_{AuNP\text{-}LYZ}^{pri}(°)$ |
|---|---|---|---|---|---|---|---|
| | | | $\Delta\varphi_{LYZ}^{sec}(°)$ | $\Delta\varphi_{AuNP\text{-}LYZ}^{sec}(°)$ | | $\Delta\varphi_{LYZ}^{pri}(°)$ | $\Delta\varphi_{AuNP\text{-}LYZ}^{pri}(°)$ |
| Measured | | 52.16 | 52.48 | | 64.96 | 64.2 | |
| | | | 0.32 | | | -0.76 | |
| $\gamma\sim30°$ | | 50.72 | | 51.6 | 69.44 | | 64.64 |
| | | | | 0.88 | | | -4.8 |
| Fitted | $N_b=0$ | 53.4 | | | 72.6 | | |
| $\gamma=29.5°$ | $N_b=120$ | | 53.8 | | | 72.4 | |
| | | | 0.4 | | | -0.2 | |
| | $N_b=16$ | | | 54.3 | | | 72.0 |
| | | | | 0.9 | | | -0.6 |
| | $N_s=0$ | 53.4 | | | 72.6 | | |
| | $N_s=120$ | | 53.7 | | | 73.3 | |
| | | | 0.3 | | | 0.7 | |
| | $N_s=16$ | | | 54.4 | | | 72.3 |
| | | | | 1.0 | | | -0.3 |
| Ideal | $N_b=0$ | - | | | 51.8 | | |
| $\gamma=28°$ | $N_b=8$ | | - | | | 51.8 | |
| | | | | | | 0.0 | |
| | $N_b=8$ | | | - | | | 51.8 |
| | | | | | | | 0.0 |
| | $N_b=0$ | 48.6 | | | 54.0 | | |
| $\gamma=33.64°$ | $N_b=8$ | | 49.0 | | | 53.8 | |
| | | | 0.4 | | | -0.2 | |
| | $N_b=8$ | | | 49.4 | | | 53.6 |
| | | | | 0.8 | | | -0.4 |
| | $N_b=0$ | 49.4 | | | - | | |
| $\gamma=38°$ | $N_b=8$ | | 49.4 | | | - | |
| | | | 0.0 | | | | |
| | $N_b=8$ | | | 49.6 | | | - |
| | | | | 0.2 | | | |

Adherence on the hillside is realistic, since this location is preferred caused by the non-topographical adhesion modulation on the laser treated surfaces according to our previous studies [18, 23-26, 50]. By supposing that the biomolecules and bioconjugates adhere at the hillsides, there are various representative locations. An interesting case is the highest hillside (x=300 nm) location of 16 AuNP-LYZ bioconjugates, where the sensitivity of the secondary peak is still enhanced, and when the first linear array of 120 LYZ biomolecules is coincident with this location. the same $N_{fitted\_chip+LYZ\_side} = 120$ LYZ shells and

$N_{fitted\_chip+AuNP-LYZ\_side} = 16$ number of AuNP-LYZ core-shells per unit cell results in $\Delta\varphi^{secondary}_{fitted\_chip+LYZ\_side} = 0.3°$ and $\Delta\varphi^{secondary}_{fitted\_chip+AuNP-LYZ\_side} = 1.0°$ shift of the secondary peaks, which shift is slightly smaller / larger than the measured values. These amounts of LYZ shells and AuNP-LYZ bioconjugates per unit cell caused $\Delta\varphi^{primary}_{fitted\_chip+LYZ\_side} = 0.7°$ and $\Delta\varphi^{primary}_{fitted\_chip+AuNP-LYZ\_side} = -0.3°$ primary peak shift, respectively.

By supposing adherence at the hillsides instead of at the bottom in the optimal azimuthal orientation, the secondary and primary peak is shifted by smaller and larger degree in case of LYZ, while larger and smaller shift is achieved after AuNP-LYZ seeding, respectively. However, the larger forward shift of the primary peak in case of hillside LYZ adherence is accompanied by a noticeable broadening, revealing that the secondary peak monitoring is more suitable for biodetection. In contrast, important advantage of the secondary peak that its smaller FWHM ensures larger FOM, which is usually defined as the resonance peak shift / (refractive index unit · FWHM).

Explanation of the enhanced sensitivity at the secondary minima, which is further enhanced in presence of Au NPs, and the anomalous sensitivity experienced in case of hillside LYZ adherence is presented in *Section 3.3.*

*3.3. Near-field distribution on a fitted chip*

In 29.5° azimuthal orientation of the bare fitted chip the **E**-field distribution indicates a global maximum on the left side of the valleys, however at the secondary peak the global maximum is shifted with a larger extent towards the valley centers, and there is a local maximum on the right edge of the valleys as well (Fig. 3a, b/a-to-d, bottom: left and right). At the secondary peak on the vertical cross-section along the valley at the turning line the **E**-field enhancement below the metal layer proves the co-existence of a glass side plasmon as well (Fig. 3a, b/a-to-d, bottom, middle). The $E_y$ field component is antisymmetric horizontally at both peaks, however the turning line is shifted with a larger extent towards the succeeding hills at the secondary peak (Fig. 3a, b/a (x=375 nm)-to-d (x=275 nm), top: left). On the vertical cross-section along the valleys at the turning line the $E_y$ field component is antisymmetric / slightly hybrid at the secondary / primary peak (Fig. 3a, b/a-to-d top: middle). Similarly, on the vertical cross-section taken perpendicularly to the grating the $E_y$ field component is antisymmetric / hybrid at the secondary / primary peak (Fig. 3a, b/a-to-d, top: right). Comparison of the $E_y$ field distributions confirms that a LRSPP possessing a horizontally and vertically antisymmetric longitudinal component propagates along the right edge of the valleys at the secondary peak, while at the primary peak the vertically hybrid $E_y$ component reveals that a SRSPP mode exists. The long- and short-range modes possess small / large attenuation determined by the overlapping with the metal [21, 28, 39]. Accordingly, larger / smaller interaction cross-section is expected at the secondary / primary peak with the biomolecules and bioconjugates, which is capable of improving the sensitivity significantly / slightly.

However, attachment of bio-objects modifies the composition of the multilayers, shifts the peaks, and transforms the near-field distribution as well. At the reflectance minima appearing in $\gamma = 29.5°$ azimuthal orientation after seeding by LYZ shells and AuNP-LYZ bioconjugates the areas corresponding to global **E**-field maxima are similarly coincident with the left border of the valleys (Fig. 3a, b/b,e and c,f, bottom: left, right). However, at the secondary peaks the global **E**-field maxima are shifted towards the succeeding hills with a larger extent, and are accompanied by local maxima at the right edge of the valleys, similarly to bare chips (Fig. 3a, b/b,c-to-e,f, bottom: left, right). The local **E**-field enhancement is larger in presence of Au NPs for both locations (Fig. 3a, b/b,e-to-c,f, bottom).

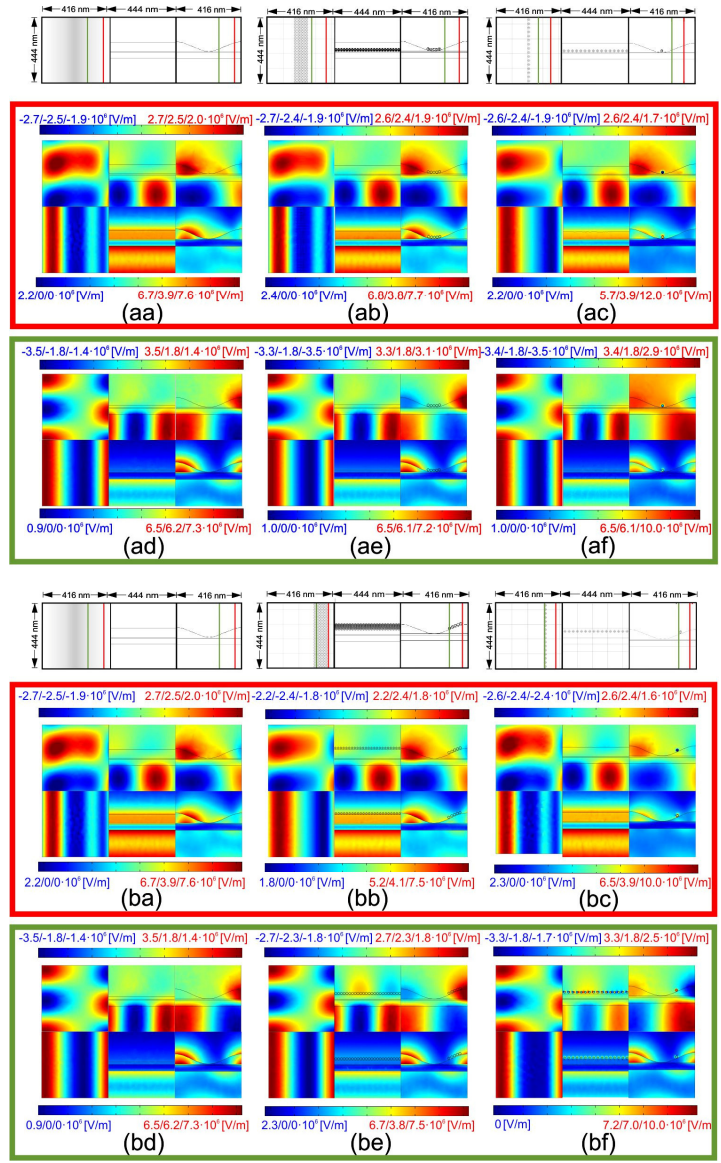

Fig. 3. The $E_y$ field component (top) and the normalized **E**-field distribution (bottom) taken horizontally (x-y plane: left) and vertically along the valleys at the turning line of $E_y$ (y-z plane: middle) and perpendicularly to the unit cell (x-z plane: right) on (a,b/ a,d) bare fitted chip, (a,b/ b,e) fitted chip covered by LYZ shells, (a,b/ c,f) fitted chip covered by AuNP-LYZ bioconjugates, at tilting corresponding to minima on the reflectance of a fitted chip in 29.5° azimuthal orientation, (a,b/ a-c) secondary and (a,b/ d-f) primary peaks in case of bio-objects at the (a/ a-f) bottom of valleys and (b/ a-f) side of the hills. The schematic drawings indicate the structure contours in different plane cross-sections and the turning lines of the $E_y$ field component at the secondary (red) and primary (green) peaks.

At the secondary peaks in case of LYZ and AuNP-LYZ covering the turning line of the antisymmetric $E_y$ longitudinal component is slightly backward shifted (x=364 nm, x=350 nm), when the bio-objects are at the bottom of the valleys, and is coincident with that on the bare chip (x=375 nm) in case of bio-object location at the side of the hills, respectively (Fig. 3a, b/b and c, top: left). At all secondary peaks the $E_y$ longitudinal component is antisymmetric on the vertical cross-sections taken along the valleys at the turning lines, as well as perpendicularly to the unit cell, for both biomolecule and bioconjugates locations (Fig. 3a, b/b and c, top: middle, right). These near-field phenomena prove that horizontally and vertically antisymmetric LRSPP modes exist at all secondary minima. The advantage of the bio-objects hillside location is that they approach the turning line of the $E_y$ component with a more well-defined antisymmetry, as well as the local normalized **E**-field maxima on the right edge of the velleys, which has potential to maximize the interaction cross-section with the LRSPP.

However, at the secondary peaks arising after LYZ covering, location of the 120 LYZ shells per unit cell at the bottom of valleys still ensures overlapping with both the global and local **E**-field maxima, while at the hillside good overlapping of multiple LYZ arrays is ensured only with the smaller local maximum on the right edge of the valleys (Fig. 3a and b/b, bottom: right). As a consequence, the secondary peak exhibits a slightly reduced sensitivity, when wide multiple arrays of 120 LYZ islands are re-located to the hillside. In contrast, in case of AuNP-LYZ, the secondary peak shift exhibits a location dependence according to the expectations, namely slightly larger sensitivity was observed in case of hillside location. This is due to that location of the 16 AuNP-LYZ core-shells at the bottom of valleys ensures compromised overlapping with both the global and local **E**-field maxima, while at the hillside better overlapping is ensured for the single AuNP-LYZ array with the local maximum corresponding to LRSPPs at the right edge of the valleys (Fig. 3a, b/c bottom: right). As a result of better overlap with a the horizontally and vertically anti-symmetric LRSPP, the secondary peak exhibits a slightly enhanced sensitivity in case of the inspected hillside location (x=300 nm) of AuNP-LYZ. Presence of AuNPs results in additional local enhancement both at the bottom of the valleys and at the side of the hills, this explains that significantly smaller amount of LYZ can be detected by monitoring the secondary peak (Fig. 3a, b/c bottom: right).

At the primary minima the turning lines of the antisymmetric $E_y$ longitudinal component are unmodified (x=275 nm) / noticeably forward shifted (x=300 nm and x=310 nm) with respect to the bare chip in presence of LYZ shells and AuNP-LYZ bioconjugates location at the bottom of valleys / side of the hills (Fig. 3a, b/e and f, top: left). For both coverings of LYZ and AuNP-LYZ the $E_y$ longitudinal component is hybrid / anti-symmetrical on the vertical cross-section taken at the turning line in case of bio-object location at the bottom of valleys / side of the hills, while it is hybrid perpendicularly to the unit cell (Fig. 3a, b/e, f top: middle, right). These near-field phenomena prove that for both coverings by LYZ shells and AuNP-LYZ bioconjugates, SRSPPs and horizontally and vertically antisymmetric LRSPPs exist at the primary peak in case of location at the bottom of valleys and on the hillsides, respectively.

At the primary peaks location of the 120 LYZ shells at the bottom of valleys cannot ensure overlapping with the global **E**-field maxima on either neighboring hills, while in case of hillside location overlapping is ensured with both the turning line of the antisymmetric $E_y$ at the valley edge and the global normalited **E**-field maximum on the succeeding hill (Fig. 3a and b/e, bottom: right). Accordingly, the primary peak exhibits an enhanced sensitivity, when the first line of the wide multiple arrays of 120 LYZ islands are relocated to x=300 nm on the hillside. As a result, the primary peak exhibits a reduced and a strongly enhanced sensitivity with respect to the secondary peak in case of LYZ location at the bottom of valleys and side of hills, respectively.

Location of the 16 AuNP-LYZ shells at the bottom of the valleys cannot ensure good overlapping with the global **E**-field maxima on either neighboring hills, while relocation to the hillside eliminates / does not make possible overlapping with the global **E**-field maximum on the preceding / succeeding hill (Fig. 3a and b/f, bottom: right). Accordingly, the primary peak exhibits a reduced sensitivity, when 16 AuNP-LYZ shells are relocated to x=300 nm on the hillside. As a consequence, the primary peak exhibits a backward shift, which is slightly and significantly reduced with respect to the secondary peak shift in case of AuNP-LYZ location at the bottom of valleys and side of the hills, respectively. The sensitivity is smaller in case of the primary peak in accordance with absence / despite the presence of LRSPPs, respectively.

At hillside location the 120 LYZ shells are moved from / into the **E**-field maxima on the preceding / succeeding hills at the secondary / primary peak, which causes decreased / increased sensitivity compared to that achievable, when the biomolecules are at the bottom of the valleys. The 16 AuNP-LYZ bioconjugates at hillside locations are closer to the turning lines of $E_y$ component, and properly overlap / do not coincide with the local / global **E**-field maxima on the succeeding hill at the secondary / primary peak. As a consequence, the secondary / primary peak exhibits larger / smaller sensitivity in case of hillside location of the AuNP-LYZ bioconjugates.

### 3.4. Reflectance originating from RGC-SPR on designed chip

To prove that the enhanced sensitivity experienced in presence of horizontally and vertically antisymmetric LRSPPs is a general phenomenon, and a maximal enhancement is achievable in an optimal configuration, inspection of a designed chip with a minimal modulation depth corresponding to the condition of rotated grating-coupling has been also performed.

The numerical computations on the designed chip have shown that the shift of the resonance peaks caused by a specific biocovering sensitively depends on the azimuthal orientation. Namely, at $\gamma_{designed\_chip} = 28°$ azimuthal orientation only one single primary peak is observable on the reflectance at $\varphi_{designed\_chip\_\gamma=28°}^{primary} = 51.8°$, which is almost insensitive to the biomolecule seeding through $N_{designed\_chip+LYZ/AuNP-LYZ\_bottom} = 8$ number of bio-objects per unit cell, independently of the presence or absence of Au NPs (Fig. 2e).

In $\gamma_{designed\_chip} = 33.64°$ azimuthal orientation the reflectance exhibits double peaks with the same depth on all of the bare, LYZ and AuNP-LYZ covered designed chips (Fig. 2d). On the bare designed chip the secondary peak appears at $\varphi_{designed\_chip\_\gamma=33.64°}^{secondary} = 48.6°$, which is smaller than the polar angle corresponding to the measured secondary peak, according to the significantly smaller average PC layer thickness. As a result of covering the valleys by $N_{designed\_chip+LYS\_bottom} = 8$ protein shells per unit cell at their bottoms, the secondary peak appears at $\varphi_{designed\_chip+LYZ\_bottom\_\gamma=33.64°}^{secondary} = 49°$, which corresponds to $\Delta\varphi_{designed\_chip+LYZ\_bottom\_\gamma=33.64°}^{secondary} = 0.4°$ polar angle shift commensurate with the measurements.

Covering the valleys by the same $N_{designed\_chip+AuNP-LYZ\_bottom} = 8$ AuNP-LYZ bioconjugates per unit cell results in a secondary peak at $\varphi_{designed\_chip+AuNP-Lys\_bottom\_\gamma=33.64°}^{secondary} = 49.4°$, corresponding to two-times larger $\Delta\varphi_{designed\_chip+AuNP-LYZ\_bottom\_\gamma=33.64°}^{secondary} = 0.8°$ shift, which approximates well the measured shift.

Comparison of the effect of LYZ shell and AuNP-LYZ bioconjugate core-shell coverings at the bottom of the valleys proves that in case of the minimal average polymer layer thickness on the designed chip 15-times and 2-times less amount of bioseeders results in the same shift, than on the fitted chip. This indicates that the interaction cross-section between the plasmonic modes that are at play is strongly enhanced already in the valleys, when the minimal layer thickness is applied. Explanation of the underlying nanophotonical phenomena is presented in the ***Section 3.5.***

In contrast, a primary peak also appears at $\varphi_{designed\_chip\_\gamma=33.64°}^{primary} = 54.0°$ on the designed chip, however this is backward shifted in case of covering by $N_{designed\_chip+LYZ/AuNP-LYZ\_bottom} = 8$ LYZ shells and AuNP-LYZ bioconjugates per unit cell.

Namely, the primary peak appears at polar angle of $\varphi^{primary}_{designed\_chip+LYZ\_bottom\_\gamma=33.64°} = 53.8°$ / $\varphi^{primary}_{designed\_chip+AuNP-LYZ\_bottom\_\gamma=33.64°} = 53.6°$, which corresponds to backward shift of $\Delta\varphi^{primary}_{designed\_chip+LYZ\_bottom\_\gamma=33.64°} = -0.2°$ / $\Delta\varphi^{primary}_{designed\_chip+AuNP-LYZ\_bottom\_\gamma=33.64°} = -0.4°$, respectively. These results indicate that the LRSPP mode, that ensures enhanced sensitivity at the secondary peak, is not at play at these primary peaks, see *Section 3.5.*

At $\gamma_{designed\_chip} = 38°$ azimuthal orientation again one single peak appears on the reflectance at $\varphi^{secondary}_{designed\_chip\_\gamma=38°} = 49.4°$, which is a secondary peak, while there is no primary peak in this orientation. This secondary peak does not shift noticeably in case of covering by $N_{designed\_chip+LYZ\_bottom} = 8$ protein shells per unit cell, while in presence of $N_{designed\_chip+AuNP-LYZ\_bottom} = 8$ bioconjugates small $\Delta\varphi^{secondary}_{designed\_chip+AuNP-LYZ\_bottom\_\gamma=38°} = 0.2°$ polar angle shift is observable resulting in a peak at $\varphi^{secondary}_{designed\_chip\_\gamma=38°+AuNP-LYZ\_bottom\_\gamma=33.64°} = 49.6°$ (Fig. 2f).

These results indicate that the course of the reflectance curve observable in case of rotated grating-coupling phenomenon as well as the sensitivity achievable by monitoring the resonance peaks strongly depends on the azimuthal orientation. An optimal azimuthal orientation exists for each biosensing chip, in which the largest shift is achievable in polar angle. Moreover, in case of the designed chip this azimuthal orientation is also a critical one, since azimuthal angle detuning results in disappearance of one of the coupled minima. In practical applications the optimal azimuthal orientation can be determined for a specific multilayer by a feedback procedure.

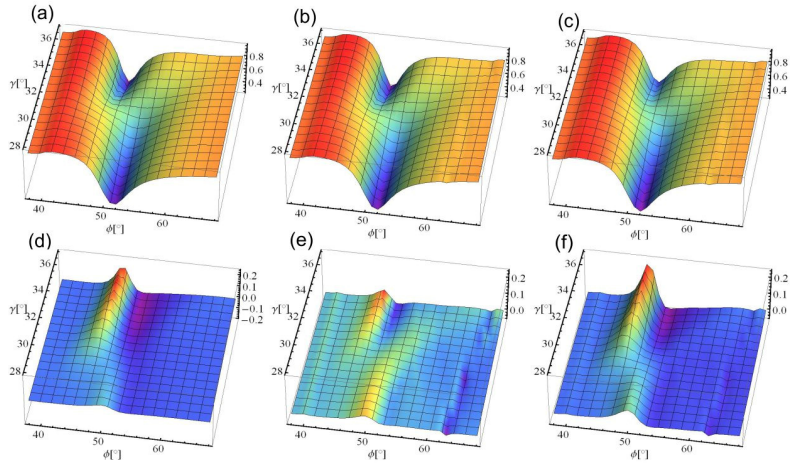

Figure 4. Calculated reflectance of different chips as a function of polar angle ($\varphi = [28°, 78°]$) and azimuthal orientation ($\gamma = [28°, 38°]$): (a) bare designed chip, (b) desigend chip covered by 8 LYZ shells per unit cell and (c) designed chip covered by 8 AuNP-LYZ bioconjugates per unit cell. (d) Difference between reflectance modifications in case of different coverings; (e, f) modifications of reflectance on a designed chip caused by seeding with (e) 8 LYZ shells and (f) 8 AuNP-LYZ bioconjugates per unit cells, compared to the bare designed chip.

In order to compare the azimuthal orientation dependent sensitivity achievable by monitoring the reflectance in case of LYZ and AuNP-LYZ seeding, we have mapped the polar and azimuthal angle dependent reflectance on the designed chip. The modification of the reflectance caused by biomolecule and bioconjugate seeding, as well as the difference between them is illustrated in Fig. 4. In case of covering either by LYZ or AuNP-LYZ bioconjugates, both the secondary and the primary minima are modified. The reflectance modification is enhanced in case of the secondary peak for both bio-coverings. Moreover, comparison of reflectance differences proves that conjugation with Au NPs results in azimuthal orientation dependent sensitivity enhancement, which effect is more significant on the secondary peak (Fig. 4d, e-to-f). This is due to the enhanced excitation of LSPR on the AuNPs via LRSPPs propagating along the valleys at the secondary peak, which is described in **Section 3.5**.

*3.5 Near-field distribution on the designed chip*

Comparison of the near-field distribution uncovers the origin of different sensitivities achievable in different azimuthal orientations of the designed RGC-SPR chip.

In $\gamma_{designed\_chip} = 28°$ azimuthal orientation the **E**-field maxima are coincident with the hills of the polymer grating (Fig. 5a/a-c, bottom: left, right). The $E_y$ component exhibits hybrid distribution horizontally as well as vertically, with a turning line at the center of the valleys (x=260 nm and 676 nm, Fig. 5 a/a-c, top). These field distributions reveal that SRSPPs exist in the valleys at the primay peak in this azimuthal orientation. Both the LYZ and the AuNP-LYZ bioconjugates are located at the **E**-field minima in the valleys, where not only the **E**-field is weak, but also the interaction length accompanying the SRSPP is small. As a consequence, the sole primary reflectance peak is not sensitive to the LYZ and AuNP-LYZ bioconjugates presence. Accordingly, no shift in polar angle is observable in this azimuthal orientation of the designed chip (Fig. 2e, 5a/a-c).

At the secondary peak appearing in $\gamma_{designed\_chip} = 33.64°$ azimuthal orientation the areas corresponding to global **E**-field maxima are coincident with the left border of the valleys (Fig. 5b/a-c, bottom: left, right). The $E_y$ longitudinal component is horizontally antisymmetric with a turning line at the right edge of the valleys (x=364 nm, Fig. 5b/a-c, top: left). In addition to this, the $E_y$ longitudinal component is antisymmetric on the vertical cross-sections at the turning line and throughout the unit cell except exactly at the valley center (where it is slightly hybrid), while perpendicularly to the grating it is hybrid (Fig. 5b/a-c, top: middle, right). These near-field phenomena prove that a horizontally and vertically antisymmetric LRSPP mode propagates along the edge of the valleys in this configuration, which ensure large interaction cross-section with the covering biolayers.

In addition to this, the **E**-field enhancement proves the co-existence of a glass side plasmon propagating along the valley (Fig. 5b/a-c, bottom: middle). The lateral extension of the **E**-field maxima ensures that both the LYZ and the AuNP-LYZ bioconjugates at the bottom of valleys are inside areas shined with an intense **E**-field originating from coupled LRSPPs (Fig. 5b/b, c, bottom). The secondary peak in reflectance is strongly sensitive to the bio-objects presence, as a result the largest polar angle shift is observable in this azimuthal orientation (Fig. 2d, 5b/a-c). This indicates that the overlap between the bio-objects and the **E**-field maxima of LRSPPs is sufficient, while the perfect overlap with the turning line of $E_y$ antisymmetry is not required to achieve sensitivity enhancement.

In contrast, at the primary peak appearing in $\gamma_{designed\_chip} = 33.64°$ azimuthal orientation smaller areas correspond to global **E**-field maxima at the left border of the valleys, which are shifted towards the valleys with a smaller extent (Fig. 5b/d-f, bottom: left, right). The $E_y$ component is antisymmetrical horizontally with a turning line at the right edge of the valleys, similarly to the secondary peak (x=364 nm, Fig. 5b/d-f, top left). In contrast, the $E_y$ component is hybrid on the vertical cross-sections at the turning line and throughout the unit cell except in the valleys, where it is symmetric, while perpendicularly to the grating it is hybrid (Fig. 5b/d-f, top: middle, right). These near-field phenomena prove that a SRSPP mode exists in the valleys and is coupled with the LRSPP at the secondary peak in this configuration. The smaller lateral extension of the **E**-field maxima allow smaller spatial overlap, and the SRSPPs can ensure inherently smaller interaction cross-section with the LYZ and the AuNP-LYZ conjugates at the bottom of the valley (Fig. 5b/e, f, bottom). As a consequence, the primary peak in reflectance is moderately sensitive to the bio-objects presence, moreover a backward shift and a slight broadening is observable caused by their adherence (Fig. 2d, 5b/d-f). Complementary studies revealed that adherence of larger amount of bio-objects can cause disappearance of this peak, according to the critical nature of this configuration.

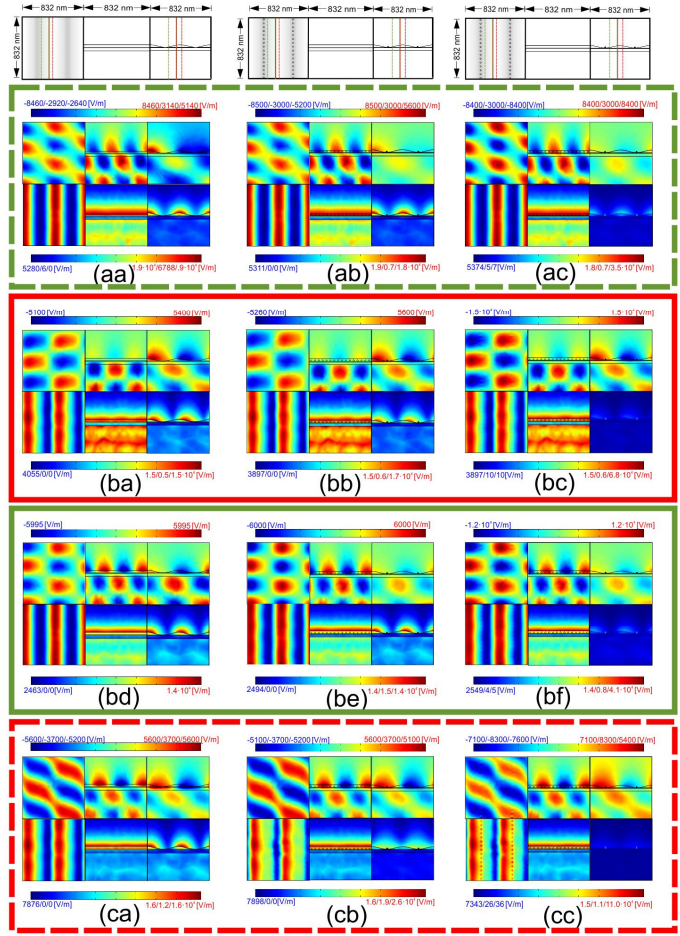

Figure 5. The $E_y$ field component (top) and normalized **E**-field (bottom) distribution in planes taken horizontally (x-y plane: left), vertically along the valleys at the turning line of the $E_y$ field component (y-z plane: middle) and perpendicularly to the unit cell (x-z plane: right), (a-c/a, b/d ) on a bare designed chip, (a-c/b, b/e) on a designed chip covered by 8 LYZ shells per unit cell, (a-c/c, b/f) on a designed chip covered by 8 AuNP-LYZ bioconjugates per unit cell, at tilting corresponding to reflectance minima, (a/a-c) in $\gamma_{designed} = 28°$ azimuthal orientation at the primary peak, (b/a-c and d-f) in $\gamma_{designed} = 33.64°$ azimuthal orientation at the secondary and primary peak, and (d/a-c) in $\gamma_{designed} = 38°$ azimuthal orientation at the secondary peak (see Visualization 1). The schematic drawings indicate the structure contours in different plane cross-sections and the turning lines of the $E_y$ field component at the secondary (red) and primary (green) peaks.

Finally, in the $\gamma_{designed\_chip} = 38°$ azimuthal orientation the **E**-field maxima cover completely the left side of the valleys (Fig. 5c/a-c, bottom: left and right). The $E_y$ longitudinal component is hybrid horizontally as well as vertically at the turning lines at the top if the hills and throughout the unit cell, as well as perpendicularly to the grating (x=416 nm and 832nm, Fig. 5c/a-c, top). These field distributions indicate that SRSPP modes exist in the valleys in this configuration. The location of both the LYZ and AuNP-LYZ bioconjugates at the bottom of the valleys is inside the **E**-field intensity maxima (Fig. 5c/b, c bottom). However, the SRSPP in the valleys possesses a reduced interaction cross-section with the covering biolayers caused by the inherently short propagation distance. As a consequence, the sensitivity of the secondary reflectance peak is moderate in this azimuthal orientation.

The coupled horizontally and vertically antisymmetric LRSPP modes excited at the secondary peak in $\gamma_{designed\_chip} = 33.64°$ azimuthal orientation of the designed chip are unique. The **E**-field distribution and large propagation distance of LRSPPs makes possible sensitivity enhancement via enhanced interaction cross-section with the LYZ shells and AuNP-LYZ bioconjugates, even if they are located at the bottom of the valleys, not exactly at the turning line of the $E_y$ antisymmetry at the valley edge. The horizontally and vertically antisymmetric LRSPPs are secondary modes, which originate from the intense laterally Bragg scattered SPPs, accordingly their wave vector equals to the projection of the original SPP modes [23]. Comparison of the **E**-field and $E_y$ distributions proves that the right edge of the valley can be considered as the metal-dielectric interface of their propagation. The horizontal and vertical antisymmetry of the $E_y$ component ensures minimal attenuation of these modes. They posses larger wavelength and are accompanied by larger propagation length (Fig. 5b/a-c). The horizontal antisymmetry also holds for the modes coupled at the primary peak, however they are accompanied by hybrid $E_y$ distribution in the vertical plane cross-section at the turning line, as a consequence they suffer larger attenuation.

The resonant excitation of AuNPs in the bioconjugates is promoted both at $\gamma_{designed\_chip} = 33.64°$ and $\gamma_{designed\_chip} = 38°$ azimuthal angles, which results in further enhanced sensitivity. Although, the local **E**-field enhancement is the largest in $\gamma_{designed\_chip} = 38°$ azimuthal orientation, i.e. when the **E**-field maxima overlap better with the AuNP arrays, the SRSPPs suffer larger attenuation in the valleys. As a consequence, larger sensitivity enhancement was achieved via the secondary peak in $\gamma_{designed\_chip} = 33.64°$ azimuthal orientation, due to the enhanced interaction cross-section with the unique LRSPPs (Visualization 1 corresponding to Fig. 5 about the $E_y$ field distribution at the secondary and primary peaks in different azimuthal orientations is provided in Supporting Information).

### 3.6 Dispersion characteristics of the modes accompanying RGC-SPR

The dispersion characteristics of the designed chip was mapped in $\gamma_{designed\_chip} = 33.64°$ azimuthal orientation and shows that two anti-crossing plasmonic bands coexist (Fig. 6). The lower branch is similar to the usual unperturbed plasmonic band on a flat dielectric-metal interface, however, it is deformed caused by the splitting originating from rotated grating-coupling. The hybridized modes exhibit a well defined turning line in their horizontally antisymmetrical $E_y$ components, which is in accordance with the phase map formed during Bragg scattering in the optimal azimuthal orientation. However, they are antisymmetric and hybrid on the vertical plane cross-sections taken at the turning lines of the $E_y$ antisymmetry along the edges of the valleys for the secondary and primary peak, respectively (Fig. 5). On the designed chip at the secondary peak a unique rotated grating-coupled LRSPP dominates due to the co-existent horizontal and vertical antisymmetry. This LRSPP is strongly coupled with a SRSPP mode at the primary peak.

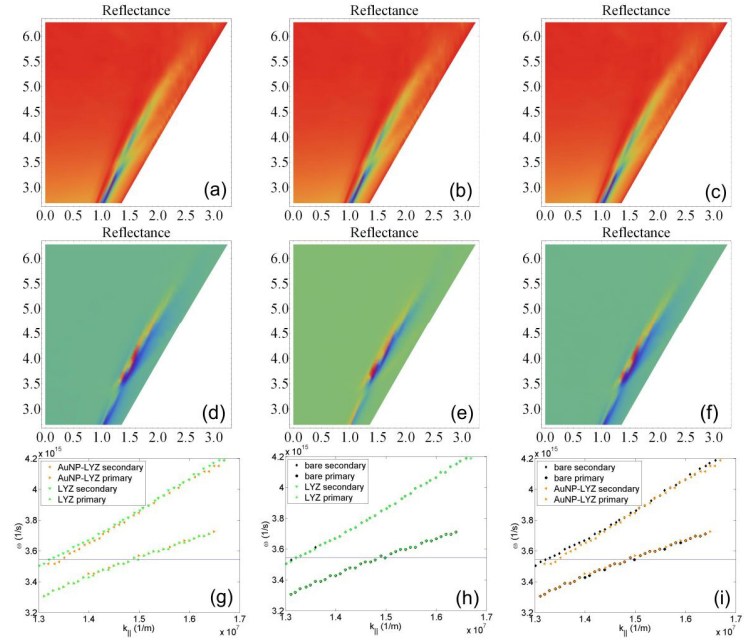

Figure 6. Dispersion characteristics in reflectance of (a) a bare designed chip, (b) a designed chip covered by 8 LYZ shells per unit cell, (c) a designed chip covered by 8 AuNP-LYZ bioconjugates per unit cell, in $\gamma_{designed} = 33.64°$ azimuthal orientation. (d) Difference between reflectance modifications, (e, f) modifications of reflectance on a designed chip caused by seeding with (e) 8 LYZ shells per unit cell and (f) 8 AuNP-LYZ bioconjugates per unit cell compared to the bare designed chip. Comparison of secondary and primary peak locations (g) 8 AuNP-LYZ to LYZ covering, (h/i) 8 LYZ / 8 AuNP-LYZ covered chip to bare designed chip

The differences in the achieved reflectances (Fig 6d-f) and the location of the reflectance minima unambiguously show that the secondary peak exhibits a significantly larger shift throughout a wide spectral interval, the largest shift is achievable at ~532 nm for the designed biochip, and the Au NPs enhanced the polar angle shift of the resonance peaks (Fig. 6 g-i).

## Conclusion

Our study revealed that there is a definite maximum on the absorption spectrum of AuNP-LYZ nanodispersion with $m_{Au}:m_{LYZ}=1:5$ close to 532 nm wavelength. Accordingly, the SPR measurements were performed on this dispersion and the numerical computations were realized for seedinga by LYZ shells and AuNP-LYZ bioconjugates with this mass ratio. The SPR measurements performed in conical mounting proved that monitoring the shift of the secondary peaks, which originate from rotated grating-coupling in the optimal azimuthal orientation, is more suitable to detect adhered bio-objects, than the observation of the broader primary peaks shift. These results are proven via FEM computations on a fitted and a designed biochip.

Numerical computations on the fitted chip revealed that the measured secondary peak shifts can be reproduced by seeding significantly larger number of protein shells and smaller number of bioconjugates at the bottom of the valleys. Comparative studies on the effect of bio-objects prove that the secondary peak is more strongly shifted than the primary one, except in case of LYZ location above certain height on the hillsides. The sensitivity can be improved by ensuring that the location of biomolecules approaches the turning line of the longitudinal $E_y$ field component at the right edge of the valleys, where the horizontally and vertically antisymmetric coupled LRSSPs propagate. The optimal location ensures overlapping with the global/local **E**-field maximum on the left / right side of the valley as well, which originate from the coupled SPPs. The enhanced / moderate sensitivity of the resonance peaks is the consequence of the large / small interaction cross-section achievable via long- / short-range modes propagating at the edges of the valleys.

The near-field study performed on the designed chip demonstrated that the **E**-field is confined onto the hills / at the edge of the hills and valleys / and onto the left side of the valleys of the polymer grating, when no primary peak shift / maximal secondary peak shift / moderate secondary peak shift is observable. These **E**-field distributions are accompanied by $E_y$ field-component distribution, which is hybrid / antisymmetrical vertically as well as horizontally with respect to the valley edge / hybrid. This proves that short-range / horizontally and vertically antisymmetric long-range / short-range modes exist in these configurations in the valleys, where the bio-objects are adhered.

It was also demonstrated that the **E**-field confinement is further enhanced in presence of Au NPs. As a result, Au NPs enhanced the difference between the polar angles corresponding to reflectance peaks on bare and bio-objects covered chips, which reveals that these particles can be used to enhance detection sensitivity. This enhancement is more pronounced for the secondary peak in the optimal azimuthal orientation.

In conclusion, considerable sensitivity enhancement is achievable in the optimal RGC-SPR configuration of the designed chip due to the secondary plasmonic modes originating from Bragg scattered of SPPs, which is further enhanced, when bioconjugates are detected. Each biosensing multilayer consisting of a wavelength-scaled grating has its own optimal configuration, which can be qualified by the optimal azimuthal orientation, capable of resulting in the largest polar angle shift. The optimal configuration is, in which the coupling efficiency of horizontally and vertically antisymmetric LRSPPS, as well as the spatial overlap between bio-objects and the LRSPPs is maximized.

**Associated contents**

**Supporting Information**

Visualization 1: Modification of the Ey component at the secondary (left) and primary (right) peak in different azimuhal orientations on horizontal (top) and vertical cross-sections (bottom).

**Funding Sources**

The research was supported by National Research, Development and Innovation Office-NKFIH through project "Optimized nanoplasmonics" K116362 and "Synthesis, structural and thermodynamic characterization of nanohybrid systems at solid-liquid interfaces" K116323. Mária Csete acknowledges that the project was supported by the János Bolyai Research Scholarship of the Hungarian Academy of Sciences.

**Acknowledgement**:

The authors would like to thank András Szenes, Hajnalka Milinszki and Lóránt Szabó for figures preparation.